\documentclass[12pt]{iopart}

\usepackage{amssymb}
\usepackage{cite}

\usepackage{bm}
\usepackage{upgreek}
\usepackage{graphicx}

\newcommand{\be}{\begin{equation}}
\newcommand{\ee}{\end{equation}}

\begin{document}

\title[]{Temperature dependence of the Casimir force}

\author{Iver Brevik$^1$ and Johan S H{\o}ye$^2$ }

\address{$^1$Department of Energy and Process Engineering,  Norwegian University of Science and Technology, N-7491 Trondheim, Norway.}
\address{$^2$Department of Physics, Norwegian University of Science and Technology, N-7491 Trondheim, Norway.}
\ead{iver.h.brevik@ntnu.no}
\ead{johan.hoye@ntnu.no}


\begin{abstract}

The Casimir force - at first a rather unexpected consequence of quantum electrodynamics - was discovered by Hendrik Casimir in Eindhoven in 1948. It predicts that two uncharged metal plates experience an attractive force because of the zero-point fluctuations of the electromagnetic field.  The idea was tested experimentally in the 1950's and 1960's, but the results were not so accurate that one could make a definite conclusion regarding the existence of the effect. Evgeny Lifshitz expanded the theory in 1955 so as to deal with general dielectric media. Much experimental work has later been done to test the theory's predictions, especially with regards to the temperature dependence of the effect.  The existence of the effect itself was verified beyond doubt by Sabisky and Anderson in 1973. Another quarter century had to pass before Lamoreaux and collaborators were able to confirm - or at least make plausible - the temperature dependence predicted by Lifshitz formula in combination with reasonable input data for the material's dispersive properties. The situation is not yet clear-cut, however; there are recent experiments indicating results in disagreement with those of Lamoreaux. In the present paper a brief review is given of the status of this research field.

\end{abstract}

\begin{center}
\end{center}

\maketitle

\section{Introduction}\label{sec:intro}

Let us begin by citing Hendrik B. G. Casimir himself:

\bigskip

"Inside a metal there are forces of cohesion and if you take two metal plates and press them together these forces of cohesion begin to act. On the other hand you can start from one piece and split it. Then you have first to break chemical bonds and next to overcome van der Waals forces of classical type, and if you separate the two pieces even further there remains a curious little tail. The Casimir force, sit venia verbo, is the last but also the most elegant trace of cohesion energy".

\bigskip
This extract is taken from Casimir's modestly formulated introductory talk at the Fourth Workshop on Quantum Field Theory under the Influence of External Conditions (QFEXT98), held in Leipzig in September 1998 \cite{casimir98}. One of us was fortunate enough to be attending this remarkable event. Casimir was then almost 90 years old, and the workshop quite appropriately took the opportunity to celebrate the 50th years' anniversary of Casimir's pioneering paper published in 1948 \cite{casimir48}. The last-mentioned paper gave a very simple derivation based upon quantum electrodynamics of how the attractive force between two neutral parallel metal plates at small separation can be envisaged as a result of the change of electromagnetic field energy in the region between the plates.  Casimir used to emphasize that the very idea of linking electromagnetic field energy to mechanical forces, in principle observable, was brought up during a conversation with Niels Bohr in Copenhagen in 1946 or 1947: After Casimir had told Bohr about his latest works on van der Waals forces, Bohr thought this over, and then mumbled something like "this must have something to do with zero-point energy". That was all, but in retrospect Casimir said he owed much to this remark.

So, in a strict sense and in conformity with the above statement of Casimir in his introductory talk, one might say that the Casimir effect concerns the case of relatively large separations between plates only (called the "tail" above), where the so-called retardation effects due to the finite velocity of light play a role. If that  view were to be upheld, the practical importance of the Casimir effect would be rather limited. In common usage the Casimir effect has however been taken to mean  also  cases where the separations between media are small. That means, one also incorporates   situations in which retardation effects  are unimportant. The latter class of phenomena goes traditionally under the name of van der Waals forces. Thus, Casimir forces and van der Waals forces are concepts used for the most part interchangeably nowadays. As is known, these kind of forces are the dominant interactions between neutral particles on nanometer to micrometer length scales. This makes the effects ubiquitous in physics, chemistry, and also biology. The effects are encountered, for instance, in the action of detergents, in the self-assembly of viruses, and even in the abilities of geckoes to climb flat surfaces.

Elementary introductions to the Casimir effect can be found in many books on quantum mechanics and quantum electrodynamics, for instance, that of Power \cite{power63}. Readers interested in more advanced treatises may consult books of Bordag {\it et al.} \cite{bordag09}, of Milton \cite{milton01}, or also extensive review articles of Milton \cite{milton04} and of Plunien {\it et al.} \cite{plunien86}. A nice presentation of regularization schemes for the Casimir effect was given by Reuter and Dittrich in this journal \cite{reuter85}.

Whereas the main properties of the Casimir effect are well known by now, there are issues related to the {\it temperature dependence} of the effect that are still insufficiently understood and subject to lively discussion in the contemporary literature. We have therefore found it useful to give a brief review of the state of art in this field,  at a level that we think should  be accessible for general physicists as well as for graduate students. Also, undergraduates ought to be able to follow  the essentials from our presentation below.


\section{Basic theory. The Lifshitz formula}\label{sec:theory}

The typical Casimir setup is illustrated in Fig.~1;  two parallel metal plates are separated by a gap of width $a$. We shall assume the plates to be nonmagnetic.
\begin{figure}[htbp]
\includegraphics[width=2in]{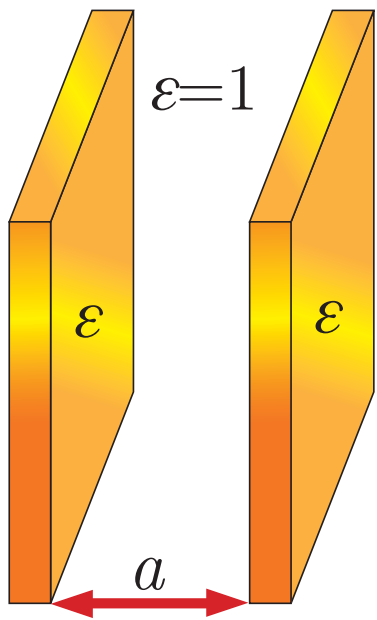}
\caption{Definition sketch: Vacuum gap of width $a$ between two nonmagnetic plates with permittivity $\varepsilon$ where metals correspond to
 $\varepsilon\rightarrow\infty$.}
\label{3layers.eps}
\end{figure}
The electric field between the plates must satisfy the boundary conditions, saying that the electric field component parallel to the metal surfaces is equal to zero. This implies that the electromagnetic field will have discrete eigenfrequencies analogous to the eigenfrequencies of a violin string. These oscillations are quantized as harmonic oscillators which have ground state energy $\hbar\omega/2$ where $\omega$ is the angular frequency and $\hbar=h/(2\pi)$ the reduced Planck's constant. The difference in ground state energy between the cases where $a$ is infinite and where $a$ is finite leads to an attractive force between the plates. This is the Casimir force. Casimir found that the force per unit surface  area (the pressure) for metals at zero temperature is
\begin{equation}
f_c=-\frac{\pi^2\hbar c}{240 a^4}
\label{1}
\end{equation}
(negative sign means an attractive force).

The Casimir force has turned out to be quite difficult to measure experimentally. This is understandable, all the time that the force is small - according to equation (\ref{1}) only  1.3 mPa (about  $10^{-8}$ atmospheres) when the separation is $a=1~\mu$m.

In 1955 Lifshitz \cite{lifshitz55} derived a more general expression for the Casimir force between two identical parallel dielectric plates of (relative) permittivity $\varepsilon$. As before, we assume the media to be nonmagnetic. The expression also holds  for finite absolute temperatures $T$. The expression is relatively complicated,

\begin{equation}
\label{2}f_c=-\frac{k_BT}{\pi}{{\sum_{m=0}^\infty}\,\vspace*{1mm}'} \int\limits_{\zeta_m}^\infty q^2\,dq \left[\frac{A_m e^{-2qa}}{1-A_m e^{-2qa}}+ \frac{B_m e^{-2qa}}{1- B_m e^{-2qa}} \right]. \label{2}
\end{equation}
 Here   $\zeta_m=   2\pi mk_B T/\hbar$ with  $m$ integer  are called the Matsubara frequencies, and play an important role when $T$ is finite. The $\zeta_m=-i\omega_m$ (or $\zeta_m=i\omega_m$, dependent on which convention is used) are caused to be discrete because of quantum mechanics.
 Moreover,  $c$ is the velocity of light in vacuum, $k_B$ is Boltzmann's constant, and the prime on the summation sign means that the term with $m=0$ is to be taken with half weight. The constants $A_m$ and $B_m$ are defined by
 \begin{equation}
A_m=\left(\frac{\varepsilon p-s}{\varepsilon p+s}\right)^2, \quad
B_m=\left(\frac{ s-p}{s+p}\right)^2,
\label{3}
\end{equation}
with
\begin{equation}
s^2=\varepsilon-1+p^2, \quad p=\frac{qc}{\zeta_m}.
\label{4}
\end{equation}
Expression (\ref{2}) makes use of imaginary frequencies.  As the physical frequencies are $\omega_m=i\zeta_m$,   the permittivity $\varepsilon$ is represented as $\varepsilon=\varepsilon(i\zeta_m)$. The coefficients $A_m$ and $B_m$ are  the squares of the reflection coefficients for respectively the TM (transverse magnetic) and the TE (transverse electric) electromagnetic waves in the region between the plates. In each of the two cases either the magnetic field, or the electric field, are parallel to the plates.
 Expression (\ref{1}) holds for all densities of the material, and for all $T \geq 0$.

From the expression (\ref{2}) it is seen that the temperature occurs at the following three places:

$\bullet$ in the prefactor of the sum;

$\bullet$ in the lower limit $\zeta_m$ of the integral, and

$\bullet$ in a possible temperature  dependence of the dissipation parameter $\nu$; see equation (\ref{12}) below.

\bigskip

We shall in the following consider  the case of metals.  This has conventionally been taken to mean that $\varepsilon(i\zeta_m)\rightarrow \infty$. At first one might think that this case is unproblematic: simply plug in the appropriate value of $\varepsilon$ and calculate the sum and integral in (\ref{2}) analytically or numerically. However, here one encounters a mathematically delicate problem, for the TE mode in the limit of zero frequency. Specifically,
\begin{equation}
 {\rm if~first~} \varepsilon \rightarrow \infty~{\rm and~then~} \zeta \rightarrow 0,~{\rm the~} B_0\rightarrow 1,~\rm while \label{5}
\end{equation}
\begin{equation}
{\rm if~first~} \zeta \rightarrow 0~{\rm and ~thereafter~} \varepsilon \rightarrow \infty,~{\rm the~} B_0 \rightarrow 0.
\label{6}
\end{equation}
These two options have given rise to two different models of a metal, namely the Ideal Metal-model (IM), and the Modified Ideal Metal model (MIM). We will now consider these two cases more closely.

\section{Ideal metals}

This served as  the standard model for the Casimir effect  of metals for several years. The IM model  was introduced in the classic paper of Schwinger {\it et al.} from 1978 \cite{schwinger78}. It  follows  option (\ref{5}) above, and means  that the contributions from the TE and TM modes are  equal to each other, including the case of  zero frequency,
\begin{equation}
A_m=B_m=1, \quad m=0,1,2,3,...
\label{7}
\end{equation}
Mathematically,  the integral in equation (\ref{2}) can be calculated analytically when $T=0$. Here we  give the results for $T=0$, and include the corrections for low temperatures. The Casimir force takes the form
\begin{equation}
f_c=-\frac{\pi^2\hbar c}{240 a^4}\left[ 1+\frac{1}{3}\left( \frac{2k_BTa}{\hbar c}\right)^4 \right], \quad \frac{ak_BT}{\hbar c} \ll 1.
\label{8}
\end{equation}
For the free energy $F$ per unit surface, determined by $f_c=-\partial F/\partial a$, the result is
\begin{equation}
F=-\frac{\pi^2\hbar c}{720a^3}\left[ 1+  \frac{45\zeta(3)}{\pi^3}  \left(\frac{2ak_BT}{\hbar c}\right)^3-\left(\frac{2ak_BT}{\hbar c}\right)^4 \right], \quad \frac{ak_BT}{\hbar c} \ll 1, \label{9}
\end{equation}
where $\zeta(3)$ means the Riemann zeta-function with 3 as argument. (Actually the middle term in this equation, independent of $a$, requires  separate attention; cf., for instance,  Refs.~\cite{milton01} or \cite{hoye03}.)

Finally, we shall be interested in the entropy $S$, which is given by the thermodynamic relation $S=-\partial F/\partial T$. We get
\begin{equation}
S=\frac{3k_B\zeta(3)}{2\pi}\left( \frac{k_BT}{\hbar c}\right)^2   -\frac{4k_B\pi^2a}{45}\left( \frac{k_BT}{\hbar c}\right)^3, \quad \frac{ak_BT}{\hbar c} \ll 1. \label{11}
\end{equation}
From this it is seen that $S=0$ when $T=0$. This is Nernst's theorem, also called the third law of thermodynamics. (Actually it is stated  more correctly  by saying that  $S=$ constant, independent of other parameters, at $T=0$.) This theorem will be a central point in the present discussion. The IM model thus   satisfies this basic requirement from thermodynamics right away.

\section{Modified ideal model, and its generalizations}

In view of the satisfactory behavior of the IM model noted above, one may ask: Why should there be any reason for changing this model at all? A problem is that a real material has to satisfy a realistic dispersion relation. In practice, the following dispersion relation, called the Drude relation, is followed by metals to a reasonably good accuracy,
\begin{equation}
\varepsilon(i\zeta)=1+\frac{\omega_p^2}{\zeta(\zeta+\nu)}.
\label{12}
\end{equation}
Here $\omega_p$ is the plasma frequency, and $\nu$ is the dissipation parameter (describing ohmic resistance in the metal). In all real metals, $\nu$ stays finite,  this being related to impurities which are always present. It turns out that the Drude relation very accurately fits optical experimental data for $\zeta <2\times 10^{15}$ rad s$^{-1}$ \cite{lambrecht00,lambrecht00a}. A typical example is gold, for which $\omega_p=9.03$ eV, $\nu=0.0345$ eV. For the Drude model, or more generally whenever
\begin{equation}
\lim_{\zeta \rightarrow 0}\zeta^2[\varepsilon(i\zeta)-1)=0, \label{13}
\end{equation}
the  zero-frequency TE mode does not contribute to the Casimir force. The first to emphasize this kind of behavior were Bostr{\"o}m and Sernelius \cite{bostrom00}, and detailed discussions were given in  \cite{hoye03} and \cite{brevik08}. There are several other papers arguing along similar lines. From a different viewpoint Jancovici and Samaj
\cite{jancovici05} and Buenzli and Martin \cite{buenzli05} considered a classical plasma of free charges in the high-temperature limit, and found the linear dependence in $T$ in the Casimir force to be reduced by a factor of 2 from the IM model prediction.

According to the information coming from the dispersion relation we thus ought to use
\begin{equation}
A_0=1, \quad B_0=0, \label{14}
\end{equation}
as input values in the Lifshitz formula (\ref{2}). At first sight the above equations (\ref{8}), (\ref{9}) and (\ref{11}) are then replaced by
\begin{equation}
f_c=-\frac{\pi^2\hbar c}{240 a^4}\left[ 1+\frac{1}{3}\left( \frac{2ak_BT}{\hbar c}\right)^4\right] +\frac{k_BT}{8\pi a^3}\zeta(3),  \label{15}
\end{equation}
\begin{equation}
F=-\frac{\pi^2\hbar c}{720 a^3}\left[1+\frac{45 \zeta(3)}{\pi^3}\left( \frac{2ak_BT}{\hbar c}\right)^3-
\left(\frac{2ak_BT}{\hbar c}\right)^4\right]
 +\frac{k_BT}{16\pi a^2}\zeta(3), \label{16}
\end{equation}
\begin{equation}
S=\frac{3k_B\zeta(3)}{2\pi}\left( \frac{k_BT}{\hbar c}\right)^2 -\frac{4k_B\pi^2a}{45}\left(\frac{k_BT}{\hbar c}\right)^3   -\frac{k_B\zeta(3)}{16\pi a^2}. \label{17}
\end{equation}
The most striking property of these expressions is that $S(0)= -k_B\zeta(3)/(16\pi a^2)$, thus violating Nernst's theorem.

 A lively discussion on this point has taken place in the literature. Arguments have even been given to give up the Drude dispersion model as such and  replace it with the "plasma model" which effectively means setting the dissipation parameter $\nu$ in (\ref{12}) equal to zero. Discussions along these lines can be found, for instance, in \cite{decca05}. Like many other researchers we think, however, that such changes of the electrodynamic theory of media should be avoided. Rather, more accurate calculations are needed.
 The clash with Nernst's theorem is  a consequence of over-idealized assumptions. It appears  natural to perform more accurate calculations of the expressions (\ref{15})-(\ref{17}), within the framework of the Drude model, taking into account measured values of $\varepsilon(i\zeta)$  and  $\nu$. Thus equation (\ref{12}) is used as basis, for small $\zeta$.  Such calculations were actual done, and reported in \cite{hoye03,brevik06,hoye07}, for the case of gold. Figure 2 shows the free energy versus temperature for low $T$. The linear term actually changes into a parabolic form with horizontal slope at $T=0$. The Nernst theorem is thus not broken after all.

 The story is however many-facetted. Without going into great detail here, we mention that the property of the entropy  becoming {\it negative} for small $T$ may appear disturbing. This is actually related to the circumstance that Casimir quantities represent physical subsystems only. Therefore they are not subject to the usual thermodynamic restrictions that hold for closed systems. A detailed analysis of this point, making use of a harmonic oscillator model, is given in \cite{hoye03}.

\begin{figure}[htbp]
\includegraphics[width=3.5in]{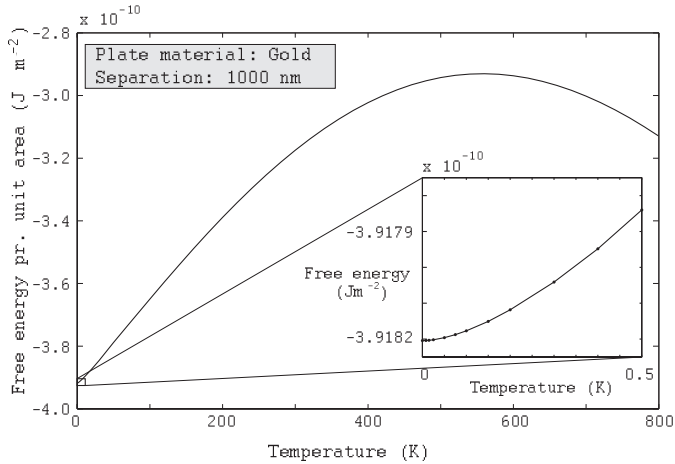}
\caption{Free energy between gold plates as a function of temperature. The inset shows the variation for small $T$. From Ref.~\cite{hoye07}.}
\label{FTfancy.eps}
\end{figure}

One special effect of the negative contribution to the entropy is that the Casimir force for metals (or more generally for large values of $\varepsilon$) decreases with increasing temperature in a certain temperature interval before it again increases to reach the classical limit for $T\rightarrow \infty$ where only the $m=0$ term in (\ref{2}) contributes. This behavior is illustrated in Fig.~3. Here the Casimir force, multiplied with the factor $a^4$ for convenience, is shown versus $a$ at constant temperature $T=300$ K. Since the force essentially depends on the product $aT$ (strictly speaking this is true for nondispersive media only), the figure effectively shows how the force varies versus $T$ for a fixed value of $a$. The approximately linear decrease between 1 and 3 $\mu$m is clearly shown, as is the linearly increasing curve for $a> 4$ $\mu$m. (The large deviation from the linear behavior below $1~\mu$m is due to the dispersion.)

\begin{figure}[htbp]
\includegraphics[width=3.5in]{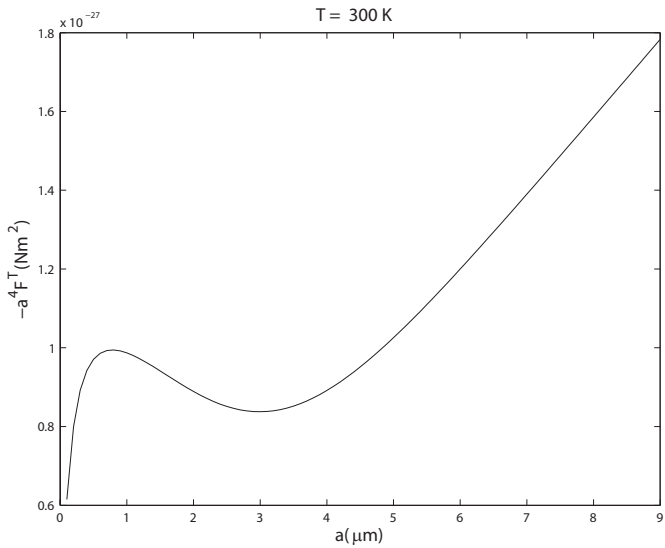}
\caption{Casimir pressure for gold plates, multiplied with  $a^4$, as function of  $a$ when $T=$ 300 K. From Ref.~\cite{hoye03}.}
\label{Fig4.eps}
\end{figure}

\section{On experiments}

As mentioned above, the Casimir attractive force is small, in practice much smaller than the electrostatic force due to the so-called "patch potentials" on the metallic test bodies, and the influence from the latter kind of forces has to  be eliminated by calibration procedures. This  is quite a demanding task for the experimentalists. Usually one will  measure the force between a microsphere and a plane, instead of between two planes, because of the strict restriction  to geometric parallelism in the latter case.

One might think: Would it not be possible in principle to find the temperature dependence of the Casimir force simply by measuring the force at some temperature $T$ and then repeat the measurement at some other  temperature $T+\Delta T?$ However, the experimentalists tell us that this is not  possible  in practice, because of disturbances and lack of stability. So all experiments to date have  been carried out at room temperature.

We mentioned above the importance of the parameter $aT$. This means that at room temperature  measurements at large distances $a$ will be of great interest in connection with the temperature dependence of the effect as then dispersion plays a decreasing role.  The problem, of course, is that at large gap widths the force becomes much smaller than it is at a typical width of 1 $\mu$m.

The Casimir force was first definitely confirmed  for dielectrics by Sabisky and Anderson in 1973 \cite{sabisky73}. A quarter of a century later, Lamoreaux demonstrated that the Casimir theory for metal plates held true \cite{lamoreaux97}. The measurements have later been reproduced by several others. In our context a most  valuable property of  the Lamoreaux experiment is that it was carried out at large distances. Lamoreaux also was involved in the newer version of this experiment  \cite{sushkov11} (see also Milton's comments in \cite{milton11}), where distances  $a$ between 0.7 $\mu$m and 7 $\mu$m were tested. Quite remarkable, the theoretical predictions based upon the Drude model were found to agree with the observed results to a high accuracy.

If this experiment stands the test of time, it will be important as it helps us understand better the electromagnetic and thermodynamic properties of real materials as well as the underlying quantum vacuum. The need for making large subtractions because of the mentioned patch potentials implies however uncertainties in the interpretation of the data in this experiment. However there are other  experiments, like as the very accurate one  of Decca \cite{decca05} carried out at small separations, which yield results apparently in accordance with the plasma model ($\nu=0$) rather than the Drude model. The reason for this conflict between experimental results is not understood  in the community. One might  suggest that the explanation has to do with the   effect called Debye shielding, known from  solid state physics and plasma physics, which can change the effective gap with between plates from the geometrically measured  width. But people doing the experiments  tell us that such  explanations seem unlikely. Also, due to the atomic structure, surfaces are not sharply defined.
 After all, and perhaps surprisingly, we can hardly do anything else than to conclude that  the thermal Casimir effect has  managed to escape from an unambiguous   explanation for  quite a long time. One might only hope, that when the explanation eventually turns up, it will reflect some deep physical property and not merely a triviality!

\end{document}